\documentclass[12pt]{article}
\usepackage{epsfig}

\newcommand{\TeV}{ \ensuremath{\,\mathrm{TeV}} }
\newcommand{\GeV}{ \ensuremath{\,\mathrm{GeV}} }
\newcommand{\eg}{e.g.\ }
\newcommand{\ie}{i.e.\ }

\newcommand{\Ast}{\ensuremath{\mathcal{A}} }

\newcommand{\shat}{\ensuremath{\hat{s}}}
\newcommand{\that}{\ensuremath{\hat{t}}}
\newcommand{\uhat}{\ensuremath{\hat{u}}}
\makeatletter
\def\lesssim{\mathrel{\mathpalette\vereq<}}
\def\vereq#1#2{\lower3pt\vbox{\baselineskip0.5pt\lineskip0.5pt
\ialign{$\m@th#1\hfill##\hfil$\crcr#2\crcr\sim\crcr}}}
\def\gtrsim{\mathrel{\mathpalette\vereq>}}
\makeatother
\begin{document}

\begin{titlepage}
\begin{center}
{\hbox to\hsize{\hfill KEK--TH--790}} 
{\hbox to\hsize{\hfill UMD--PP--02--022}}

\bigskip
\vspace{3\baselineskip}

\textbf{\Large 
Alternative Signature of TeV Strings}

\bigskip

\bigskip

\textbf{Kin-ya Oda}\\
\smallskip

\textit{\small
Theory Group,
KEK,
Tsukuba, Ibaraki 305-0801, Japan}

\medskip
\textbf{Nobuchika Okada}\\
\smallskip
\textit{\small
Department of Physics, 
University of Maryland,
College Park, MD20742, USA}

\bigskip

{\tt kin-ya.oda@kek.jp} \\
{\tt okadan@physics.umd.edu}

\bigskip
\bigskip

\textbf{Abstract}\\
\end{center}
\noindent
In string theory, 
it is well known that 
any hard scattering amplitude inevitably suffers
exponential suppression. 
We demonstrate that, if the string scale is $M_s < 2\TeV$, 
this intrinsically stringy behavior leads to a dramatic reduction
in the QCD jet production rate with very high transverse momenta
$p_T \geq 2\TeV$ at LHC.
This suppression is sufficient 
to be observed in the first year of low-luminosity
running. 
Our prediction is based on the universal behavior of string theory, 
and therefore is qualitatively model-independent.
This signature is alternative and complementary
to conventional ones such as 
Regge resonance (or string ball/black hole) production.

\end{titlepage}


\section{Introduction}
The discovery of D-branes in string theory~\cite{Polchinski:1995mt}
has greatly enhanced the phenomenological investigation of 
the interesting possibility 
that we are living in a 4 dimensional subspace (``3-brane'') 
of a larger $D$ dimensional world (``bulk''),
which reveals itself through the gravitational interaction 
above the TeV scale~\cite{Arkani-Hamed:1998rs}.%
\footnote{The idea of the extra dimensions above the TeV scale
was proposed earlier~\cite{Antoniadis:1990ew}.}
The hierarchy
between the electroweak scale and
the 4-dimensional Planck scale 
may be solved by arranging the volume of the extra $D-4$ dimensions 
so as to lower
the fundamental gravitational scale of the D-dimensional theory down to 
${\cal O}(1 \TeV)$~\cite{Arkani-Hamed:1998rs}.  
This idea can be realized as a class of models in string theory
(see Ref.~\cite{Antoniadis:1998ig} and citations thereof),
which possess the following characteristics in general. 
The fundamental string scale $M_s$ is around TeV, 
where $M_s$ is defined as $M_s=\alpha'^{-1/2}$
with $\alpha'$ being the string tension.
The graviton corresponds to the massless mode of the closed string 
in the bulk,
while the standard model fields are realized as the massless modes
of open strings whose end-points are on the D3-brane.
(In general, the ``extra'' $D-4$ dimensions are compactified with ``large'' radius
which is $\sim$ fm--mm depending on $D$,
and the remaining $10-D$ dimensions are compactified with string length.)
In this paper, we present a model-independent prediction 
derived from the universal nature of TeV scale string theory.

When the string scale is around TeV, 
we may expect signals at the Large Hadron Collider (LHC),  
whose center-of-mass energy is designed to be $\sqrt{s}=14\TeV$.
The most direct signature of the TeV strings
would be the production of excited string states
observed as a number of 
Regge resonances~\cite{Dudas:1999gz,Accomando:1999sj,Cullen:2000ef}.
We note however that there exist other new physics 
such as technicolor theories~\cite{Farhi:1981xs} or 
preon models~\cite{Marshak:1986sw} 
which also provide similar resonance events 
such as ``techni-hadrons'' or excited states of the known fermions.  
It is not easy, in a hadron collider, 
to exclude other possible particle theories and
confirm the TeV scale string theory
only from the observation of the resonance states.

In addition, there are somewhat perplexing consequences of 
semi-classical TeV scale gravity: 
black hole might be formed and hide all the interactions
for the process at $\sqrt{\shat}\gg M_D$~\cite{Giddings:2001bu,Dimopoulos:2001hw}, 
where $\sqrt{\shat}$ and $M_D$ are 
the center-of-mass energy of the scattering partons
and the $D$-dimensional Planck scale, 
respectively.%
\footnote{
The $D$-dimensional Planck scale is given by 
$M_D\sim g_s^{-2/(D-2)}M_s$,
where $g_s$ is the string coupling constant.}
It has been suggested that 
the black hole production cross section is simply given by 
the geometrical cross section $\sigma=\pi R_S^2$, 
where $R_S$ is the Schwarzschild radius of the black hole 
whose mass is equal to $\shat$. 
This claim is based on the assumption that 
the classical hoop conjecture is valid 
for the quantum process of the partons
and thus the black hole formation occurs 
whenever the impact parameter is smaller than $R_S$.
However, this proposition based on semi-classical quantum gravity
is still being debated~\cite{Voloshin:2001vs,Giddings:2001ih,Voloshin:2001fe}.

On the other hand, it has been claimed
that black hole formation occurs 
in TeV scale string theory~\cite{Dimopoulos:2001qe} 
due to the correspondence principle for black holes and strings%
~\cite{Horowitz:1997nw,Horowitz:1998jc}.  
The picture is as follows.
As one raises the energy of the string scattering, 
highly excited string states are produced 
and tangled to form ``string balls''. 
If one raises the energy further,
the amplitude of the string ball production 
would be smoothly connected with that of the semi-classical 
black hole production at the ``correspondent point''.
Here, we note that while the correspondence principle is 
safe for the static situation (entropy), 
there is no evidence of it 
for the dynamical process (S-matrix) involving non-BPS black holes.%
\footnote{We thank M. Natsuume for bringing this point to our attention.}
The problem here is that 
the energy range which can be explored at LHC 
is not high enough~\cite{Dimopoulos:2001hw}
so that the semi-classical treatment 
of the produced black hole is less reliable. 
Therefore, even if the black hole is really produced,
deeper understanding of the correspondence principle 
is necessary to analyze the physical process in detail.

In order to complement the conventional signature 
of resonance mode (or string ball/black hole) production,
which involves the complicated arguments above, 
it is important to find an alternative.
In this paper, we employ the hard-scattering behavior
of the string amplitude. 
It is well known that any 
string amplitudes are exponentially suppressed 
in the hard scattering limit~\cite{Gross:1987kz,Gross:1988ar}.
In order to demonstrate this stringy suppression,
we examine the QCD jet production rate 
with very high transverse momentum. 
We show that, depending on the string scale, 
the rate is dramatically reduced from the standard model prediction. 
This result is the reflection of the intrinsically 
stringy structure and is distinct 
from any other new (particle) physics.

\section{Hard scattering limit of string amplitudes} 
Gross and Mende have shown that any closed string amplitudes 
exhibit a universal behavior in the hard scattering limit, i.e.,
$s\rightarrow\infty$ with fixed $t/s$~\cite{Gross:1987kz}.
Since these amplitudes are 
dominated by the tachyonic form factors, 
this behavior is independent of the type of string theory
(bosonic string, superstrings, heterotic strings etc.),
the perturbative string vacuum, and external states 
of the scattering~\cite{Gross:1988ar}. 
This argument has also been extended to open
strings~\cite{Gross:1989ge}. 
The $G$-loop amplitude was found to be exponentially suppressed%
\footnote{
When initial momenta are transverse to the D-brane, 
the hard scattering 
behavior could be modified to the power-law suppression~\cite{Barbon:1996ie}. 
In our case, initial memonta are parallel to the D-brane (`our' 3-brane), 
and we may expect the same exponentially suppressed amplitude.    
} 
as
\begin{eqnarray}
\Ast_G(s,t)\sim e^{-\alpha'sf(\theta)/(G+1)},\label{eq:gross-mende}
\end{eqnarray}
where the positive function $f(\theta)$ is given in terms of 
$\lambda=-t/s=\sin^2(\theta/2)$ by
$f(\theta)=-\lambda\log\lambda-(1-\lambda)\log(1-\lambda)$.
This amplitude reproduces 
the hard scattering limits of
both the Veneziano (open string) and Virasoro-Shapiro (closed string) amplitudes
at the tree level, $G=0$.

Although the perturbation series of Eq.~(\ref{eq:gross-mende}) is 
badly divergent, it is still possible to evaluate 
the limit of the amplitude with its leading terms resummed to all orders
by utilizing Borel transform techniques~\cite{Mende:1990wt}.
The resultant suppression is found to be milder,
but is still significant
\begin{eqnarray}
\Ast_\mathrm{MO}(s,t)\sim e^{-c\sqrt{\alpha'sf(\theta)}},\label{eq:mende-ooguri}
\end{eqnarray}
where $c$ is a factor having little angle dependence 
and is in principle calculable.%
\footnote{ 
For example, we obtain $c\simeq 1.2$ for the tachyon scattering 
in the bosonic closed string theory
when $g_s$ is set equal to the QCD coupling constant. 
}
For simplicity, we set $c=1$ in the following analysis.

In the toy model of the TeV scale string theory,
it has been shown at the tree level~\cite{Cullen:2000ef} 
that every standard model amplitude
(corresponding to the open string scattering) 
is multiplied by a \emph{common} stringy form factor
\begin{eqnarray}
\mathcal{S}(s,t)=
\frac{\Gamma(1-\alpha's)\Gamma(1-\alpha't)}{\Gamma(1-\alpha's -\alpha't)},
\end{eqnarray}
which is essentially 
the Veneziano amplitude.
A similar result with different factor $\mathcal{F}(s,t,u,m^2)$
is found for the emission of the graviton 
(or the production of Kaluza-Klein excitations) 
which corresponds to the tree level closed string 
emission~\cite{Dudas:1999gz,Cullen:2000ef}.
Here, $m$ is the mass of the Kaluza-Klein mode
which is negligible when $D>6$~\cite{Antoniadis:1998ig}.
It can be confirmed that both the hard scattering limits of 
$\mathcal{S}(s,t)$ and $\mathcal{F}(s,t,u,0)$ reproduce 
the universal behavior of Eq.~(\ref{eq:gross-mende}) with $G=0$
\cite{Cullen:2000ef}.
Therefore, it is reasonable to expect
that the hard scattering amplitude (of the standard model particles)
in TeV scale string theory is described 
by the standard model amplitude
multiplied by the resummed stringy factor 
$\mathcal{A}(s,t)=e^{-\sqrt{\alpha'sf(\theta)}}$
such that
\begin{eqnarray}
\frac{d\sigma}{dt}=\left.\frac{d\sigma}{dt}
  \right|_\mathrm{SM}|\mathcal{A}(s,t)|^2.
\label{eq:basic_formula}
\end{eqnarray}
This is the basic formula in our analysis.%
\footnote{
We can explicitly show that
$\mathcal{S}(s,t)\simeq\sqrt{2\pi\alpha's}\tan(\theta/2)\,e^{-\alpha'sf(\theta)}$
utilizing Stirling's formula.
The prefactor $\sqrt{2\pi\alpha's}\tan(\theta/2)$ 
leads to the subleading contributions in Eq.~(\ref{eq:mende-ooguri})
after the Borel resummation.
Even if we naively multiply the stringy factor 
in Eq.~(\ref{eq:basic_formula})
by this prefactor,
our final prediction (namely, deficit in the QCD-jet production rate)
is modified only by a factor $< \mathcal{O}(10)$,
and hence our conclusion is not changed.
}

In what follows,
we give several comments to clarify our arguments.
Firstly, if the known particles have further substructures 
as in preon models, 
the hard scattering amplitude will, in general, become soft 
at high energies. 
An interesting feature of string theory is that 
the stringy structure makes the hard scattering process 
much softer than any local field theory can do. 
The history of hadronic interactions tells us 
that this point played a crucial role in 
excluding the dual resonance (string) model,
while giving support for the QCD parton model (see \eg Ref.~\cite{Green:1987sp}).
Thus our result is distinct from any other new particle physics. 

Secondly, we note that a string amplitude such as 
$\mathcal{S}(s,t)$ is, in general, a rapidly varying function 
with many zeros and poles. 
The poles correspond to the existence of the resonance states. 
Although the factor $e^{-\alpha'sf(\theta)/(G+1)}$ in the hard scattering limit 
is evaluated off the 
poles,  
we can regard it as the amplitude suitably averaged 
over all zeros and poles (see \eg Ref.~\cite{Green:1987sp}). 
Indeed, we can find the same suppression factor 
even on the pole, 
if we appropriately take the width into account 
and remove the singularity.

Finally, 
there is another limit of the scattering amplitude in the string theory 
called 
Regge limit, i.e.,  $s \rightarrow \infty$ with $t$ fixed~\cite{Amati:1987wq}. 
In this limit, the amplitude behaves as 
$ \sim (\alpha's)^{\alpha't+\alpha(0)}$, 
where a constant $\alpha(0)$ is the intercept.%
\footnote{
The intercept is positive for the realistic theory having massless modes
in the mass spectrum. In particular $\alpha(0)=0$ for $\mathcal{S}(s,t)$.
}
In the argument of Ref.~\cite{Dimopoulos:2001qe},
the production cross section of the string balls  
is a monotonically increasing function of $s$, 
and is connected to that of the black hole 
at the correspondent point.  
This seems to suggest that, in the string picture, 
black hole production correspond to the contributions 
from $-\alpha(0)/\alpha'\lesssim t\lesssim 0$ in the Regge region 
since the fixed angle (hard scattering) 
amplitude is exponentially suppressed.%
\footnote{This correspondence in the Regge region 
was mentioned in the earlier work~\cite{Amati:1989tn}.}

In the black hole picture, 
all the substructures smaller than the Schwarz\-schild radius 
would be hidden and 
most particles would be emitted by 
evaporation through Hawking radiation. 
Consequently the rate of the emission of the hard quanta 
higher than the Hawking temperature 
is exponentially suppressed due to the Planck distribution  
of Hawking radiation such that $\sim e^{-E/T}$. 
From this point of view it is interesting 
that, in the string picture, 
the perturbative behavior $\sim e^{-E^2}$ in Eq.~(\ref{eq:gross-mende})
is modified to $\sim e^{-E}$ as in Eq.~(\ref{eq:mende-ooguri})
when summed up to all orders.

While all our calculations in this paper are performed 
in the string picture, 
the exponential suppression of the hard scatterings 
might be considered as a consequence 
of black hole formation,
if the correspondence principle is valid also for the 
S-matrix and black holes are really formed
as is claimed in Ref.~\cite{Dimopoulos:2001qe}.

\section{High $p_T$ jet production rate}
In order to demonstrate the stringy effect in the hard scattering limit,
we study the QCD jet production rate with very high transverse momentum.
At LHC, high $p_T$ events can be triggered using reasonable
$p_T^{\min}=\mathcal{O}(100\GeV)$~\cite{LHCcombined} 
which is far below the scale $\mathcal{O}(1\TeV)$ considered here. 
We take $p_T^{\min}=2\TeV$,
for which trigger efficiency would be practically 100\%.
The jet resolution is expected to
be of order $10\GeV$, hence negligible for our purposes.
The jet production cross section
(which is dominated by QCD scatterings among quarks and gluons)
for a given interval of the transverse momentum $[p_T,\,p_T+\Delta p_T]$
is evaluated by 
\begin{eqnarray}
\lefteqn{\Delta\sigma_{[p_T,\,p_T+\Delta p_T]}(s)=}\nonumber\\
  &&\int_0^1dx_1 \int_0^1dx_2 \int d\that\,
    \sum_{ijkl}\frac{1}{1+\delta_{kl}}\,
    f_i(x_1,Q)\,f_j(x_2,Q)    
\nonumber\\
  &&
  \times\left.\frac{d\hat{\sigma}_{ij\rightarrow kl}}{d\that}\right|_\mathrm{SM}|
    \mathcal{A}(\shat,\that)|^2,
\end{eqnarray}
where $\shat$, $\that$ and $\uhat$ are the Mandelstam variables
of the parton-parton scattering obeying $\shat=x_1x_2s$
and $\shat+\that+\uhat=0$;
the region of integration for $\that$ is
determined by the conditions 
$p_T^2<\that\uhat/\shat<(p_T+\Delta p_T)^2$
and $\shat+\that>0$ for each set of $x_1$ and $x_2$;
$f_i$ are parton distribution functions, 
$i,j,k,l$ are summed over gluon and quark flavors, and
$Q$ is the typical energy of the parton scattering defined as 
$Q^2=2\shat\that\uhat/(\shat^2+\that^2+\uhat^2)$.
$\left.d\hat{\sigma}_{ij\rightarrow kl}/d\that\right|_\mathrm{SM}$ 
are the QCD parton-level cross sections 
of the standard model (summarized \eg in Ref.~\cite{Ellis:1996qj}). 
In our analysis, we employ CTEQ5M1~\cite{Lai:1999wy} 
for the parton distribution functions. 

\begin{figure}
\begin{center}
\epsfig{file=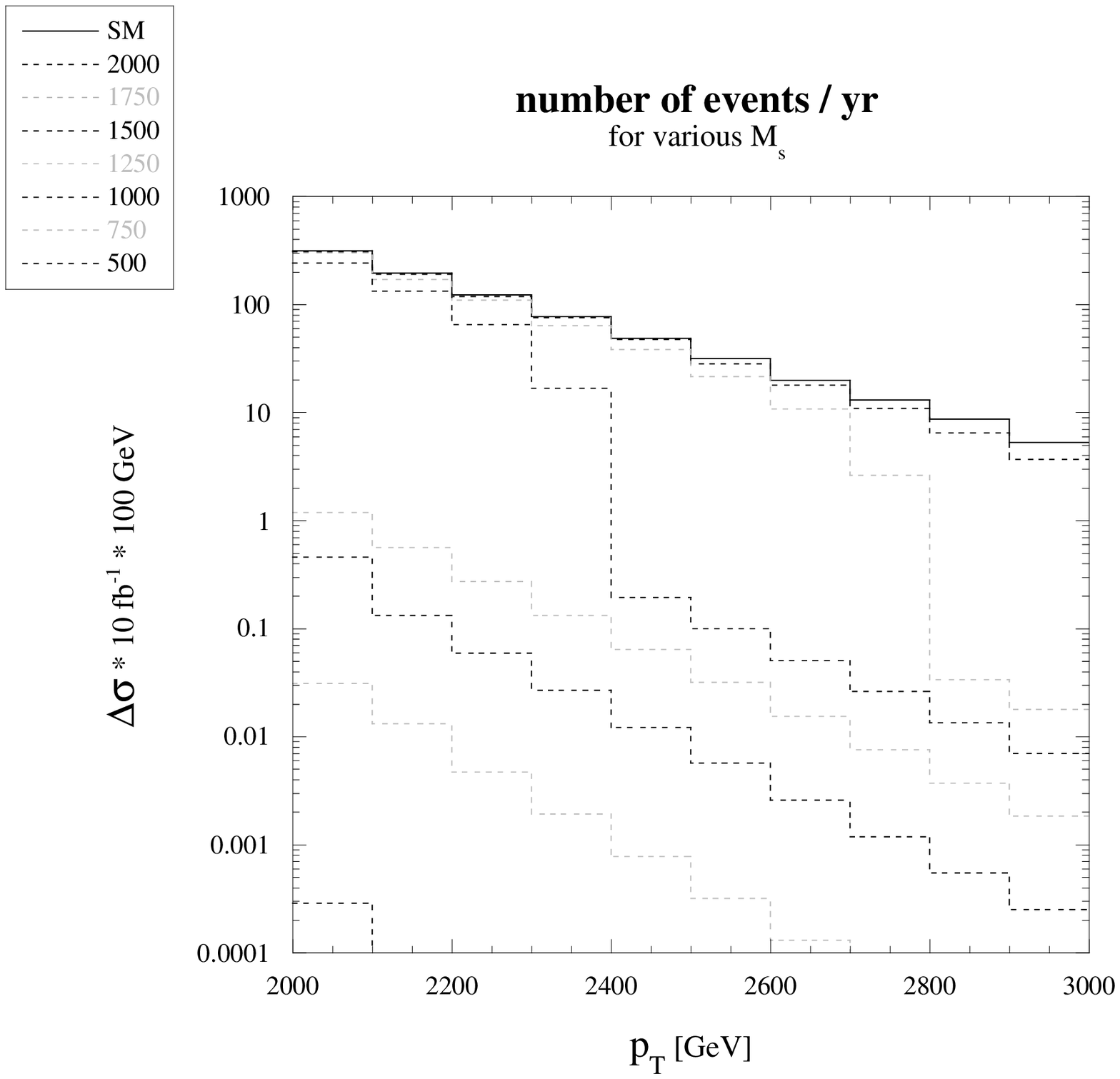,width=12cm}
\caption{Number of jet production events per year in the $100\GeV$ bin,
expected to be observed in LHC with the first
low-luminosity running ($10\,\mathrm{fb}^{-1}/\mathrm{yr}$).
The solid line is the standard model prediction and the lines below are 
the corresponding results in TeV scale string theory with the string scale
varied from $M_s=2\TeV$ to $0.5\TeV$.
} \label{fig:LHC_Ms}
\end{center}
\end{figure}

The result is shown in Fig.~\ref{fig:LHC_Ms}. 
Here, the center-of-mass energy $\sqrt{s}$ is taken 
at $\sqrt{s}=14\TeV$ as is designed for LHC. 
We have plotted 
the number of events per year for the QCD jet production
expected for the low-luminosity running ($10\,\mathrm{fb}^{-1}/\mathrm{yr}$)
within $100\GeV$ bin from $p_T=2\TeV$ to $3\TeV$.
The solid line denotes the standard model prediction and
the lines below correspond to the predictions of TeV scale string theory
with the string scale from $M_s=2\TeV$ to $0.5\TeV$.%
%
%

In this analysis, we have applied the 
stringy factor $e^{-\sqrt{\alpha'\shat f(\theta)}}$
only when $\shat > 10 M_s^2$ 
(\ie we have conservatively
applied $\mathcal{A}(\shat,\that)=1$ when $\shat \leq 10 M_s^2$).%
\footnote{
It is estimated that the hard scattering limit is applicable at 
$N\gtrsim (g_s^2/4\pi)^{-1}$,
where $N\sim\alpha'\shat$~\cite{Cornet:2001gy}.
Note that we are treating QCD scatterings 
with the coupling constant $g\sim 1$
and also that the string coupling constant
is expected to be of $\mathcal{O}(1)$
in the original consideration of 
the TeV scale string theory~\cite{Antoniadis:1998ig}.
}
The result barely depends on the value of this cutoff;
the condition $\shat > 10 M_s^2$ is already satisfied in 
the most regions of integration
that we perform at very high transverse momenta.%
\footnote{
More explicitly, 
this condition is satisfied in all the region of $2\TeV \leq p_T \leq 3\TeV$
for $M_s < 1.5\TeV$.
In Fig.~\ref{fig:LHC_Ms},
we can observe the influence of this cutoff
at $p_T<2.4\TeV$, $2.8\TeV$ and $2\TeV \leq p_T \leq 3\TeV$
for $M_s=1.5\TeV$, $1.75\TeV$ and $2\TeV$, respectively.
}

\section{Summary}
We have demonstrated that 
the exponential suppression of hard scattering amplitude 
in TeV scale string theory
leads to a dramatic reduction in the production rate 
of the QCD jets with very high transverse momentum $p_T\geq 2\TeV$ 
at LHC when the string scale is at $M_s<2\TeV$. 
This deficit is sufficient to be observed in the first year low-luminosity running.
This signature is complementary 
to the conventional signature of Regge resonance (or string ball/black hole) production.
We note that 
although we have presented the calculation only with
the leading QCD scatterings among quarks and gluons, 
\emph{any} other standard model cross sections will suffer
the same suppression due to the universal nature of 
the high energy behavior of the string theory. 
It is worth investigating various hard scattering processes 
in the TeV scale string theory. 

\bigskip
\bigskip

\noindent
\textit{Note added:}
{\small
  In our calculation, we have considered the string loop corrections. In the field theory, the infrared divergences from virtual soft gluon loops are canceled by the real soft gluon emissions from external lines and are absent when we treat a properly constructed jet cross section. This cancellation is not altered in the string theory. This can be explained in two ways. Firstly string amplitude is calculated in such a way that external lines are inserted as vertex operators, which is already mode-expanded and "field theoretical limit" is taken (by utilizing the modular invariance), and therefore treatment of this type of infrared divergences (i.e. argument of attaching virtual and real soft gluon lines onto external lines) is not modified from the field theoretical treatment. Secondly the fact that the sum of virtual and real soft gluon contributions may be calculated by a properly constructed jet cross section follows from the unitarity of the theory. We note that the theory remains unitary when we go beyond the low energy effective theory to string theory.

  The precise value of infrared cut-off is not important to our conclusion because we are estimating a kind of inclusive cross section of the high $p_T$ jet event with any number of additional jets. Though no one has ever succeeded in explicitly showing that $2 \rightarrow N$ process is always suppressed in string theory when $2 \rightarrow 2$ process is exponentially suppressed, this fact is widely believed and proving this is beyond the scope of the current paper. In this paper, we simply assume that this is the case and present that dominant $2\rightarrow 2$ contributions are suppressed.
}

\bigskip
\bigskip

\begin{center}
\textbf{Acknowledgments}
\end{center}

\noindent
We are grateful to N. Ishibashi, K. Odagiri, J. Hisano and Y. Kitazawa
for useful comments,
and A. Akeroyd and K. Odagiri for reading the manuscript.
K.O.\ would like to thank I. Kishimoto 
for helpful conversation in the early stages of this work 
and especially T. Suyama and M. Natsuume for extensive discussions.

\bibliography{paper}
\bibliographystyle{utphys}

\newpage

\end{document}